# Intégration de la synthèse mémoire dans l'outil de synthèse d'architecture GAUT Low Power

Gwenolé CORRE, Nathalie JULIEN, Eric SENN, Eric MARTIN

LESTER[1], Université de Bretagne Sud

Centre de recherches, BP 92116

56321 Lorient Cedex – France

Tel : 02/97/87/45/28 Fax : 02/97/87/45/00

e-mail : prenom.nom@univ-ubs.fr

[1] Laboratoire d'Electronique et des Systèmes TEmps Réel ; http://lester.univ-ubs.fr

**Résumé** - Les systèmes supportant des applications de traitement du signal et de l'image manipulent de plus en plus de données. Cela entraîne une utilisation intensive de la mémoire qui devient le point critique du système ; la mémoire limite les performances et représente une proportion importante de la consommation globale. Dans le cadre du développement de l'outil de synthèse d'architecture GAUT Low Power, nous nous intéressons à la synthèse de la partie mémoire. Dans la première phase de ce travail, il s'agit d'intégrer, lors de la synthèse d'architecture, les contraintes liées au mapping mémoire préalablement défini.

**Abstract** - The systems supporting signal and image applications process large amount of data. That involves an intensive use of the memory which becomes the bottleneck of systems. Memory limits performances and represents a significant proportion of total consumption. In the development high level synthesis tool called GAUT Low Power, we are interested in the synthesis of the memory unit. In this work, we integrate the data storage and data transfert to constraint the high level synthesis of the datapath's execution unit.

## 1. Introduction

L'évolution des techniques d'intégration et de la technologie permettent d'implanter des applications de plus en plus complexes. L'augmentation de la complexité et la nécessité de réduire les temps de conception (time-to-market) ont conduit au développement d'outils de synthèse d'architecture. Ces outils, généralement orientés vers la conception d'ASICs, prennent la description comportementale d'une application en entrée et fournissent une architecture de la partie traitement et contrôle au niveau RTL. Les outils de synthèse actuels s'intéressent, en priorité, à la synthèse de l'unité de traitement sans tenir compte de l'influence de la mémorisation des données. Or, les applications en traitement de signal et de l'image (TDSI) nécessitent beaucoup de ressources mémoire. La bande passante mémoire et la quantité de données manipulées entraînent une forte consommation des unités de mémorisation qui peuvent représenter de 50 à 80% de la consommation totale d'un circuit [1]. Synthétiser l'unité de traitement sans prendre en compte le placement et le transfert de données conduira à une solution mémoire peu efficace. En effet, après synthèse les possibilités d'optimisation de l'unité de mémorisation sont faibles car trop contraintes par l'unité de traitement. Il apparaît alors important, pour la synthèse d'algorithmes TDSI, d'inclure des critères d'optimisation tout au long du flot de synthèse afin d'obtenir une unité de mémorisation à coût réduit.

Dans cet article, nous proposons d'introduire une méthodologie visant à intégrer la synthèse mémoire lors de la synthèse d'architecture.

Un outil, GAUT Low Power[2] a été développé au LESTER [2] ; il intègre les critères de temps, surface et consommation pour la synthèse de l'unité de traitement. Il s'agit ici de faire évoluer cet outil en :

- intégrant une stratégie de placement des données de l'algorithme TDSI en mémoire en amont de la synthèse d'architecture.

- prenant en compte le placement des données dans les étapes de synthèse de l'outil (sélection, allocation, ordonnancement, assignation).

- intégrant la conception des générateurs d'adresses en aval de la synthèse.

Nous présenterons d'abord le contexte de l'étude en section 2 avec état de l'art sur les techniques d'optimisation mémoire à un haut niveau d'abstraction. Dans la section 3, nous explicitons la stratégie de conception mise en œuvre. Puis, la première étape du travail concernant le cœur de l'outil GAUT Low Power pour la prise en compte d'un mapping mémoire est présentée en section 4. Elle concerne essentiellement la modification de l'ordonnancement qui sera plus particulièrement détaillé ici, avant de conclure sur les perspectives et les futurs développements.

## 2. Contexte

L'optimisation de la consommation des mémoires vise d'une part à diminuer leur taille (ce qui limite la puissance statique) et d'autre part à limiter les transferts et les transitions sur les bus (ce qui réduit la puissance de commutation). Il existe différentes techniques d'optimisation dans la littérature, qui peuvent être classées suivant trois axes.

---

[2] Génération Automatique d'Unité de Traitement pour la conception d'architecture faible consommation

Une première approche s'intéresse à la conception de hiérarchie et d'architecture mémoire pour des applications dont les séquences d'accès sont connues a priori. De nombreuses méthodologies proposent des techniques permettant de sélectionner une architecture faible consommation utilisant des mémoires caches [3] ou des mémoires SRAM dédiés ou scratch-pads [4]. Le choix des architectures mémoire est effectué par rapport aux estimations en vitesse, surface et consommation obtenues à l'aide de modèles analytiques.

La seconde technique d'optimisation mémoire considère une hiérarchie et une architecture fixe ; la réduction de la consommation est réalisée par transformation du code. Ces transformations vont permettre de réduire les besoins mémoire, les séquences d'accès mémoire ou le nombre de transitions sur les adresses. Au niveau du code source, une sélection de structures de données, adaptées à l'application, permettra de réduire les besoins de mémorisation [5]. De nombreuses techniques d'ordonnancement des accès mémoire [6] et d'assignation des données en mémoire et en registres [7] ont été développées en s'orientant vers l'optimisation de la consommation. Une méthode de gestion explicite de placement dynamique des données en mémoire, développée au LESTER, permet d'optimiser les performances de la partie mémoire [8]. D'autres études [9] montrent l'apport de la limitation du nombre de transitions sur les bus d'adresse.

La troisième approche est celle adoptée par l'IMEC [10] sous la dénomination de DTSE (Data Transfer and Storage Exploration). Elle apporte de bons résultats puisqu'elle permet l'optimisation conjointe de l'architecture mémoire et du code de l'application. Cependant le temps nécessaire à l'obtention d'une solution optimale est relativement important car de nombreuses étapes ne peuvent être automatisées et requièrent l'expertise d'un concepteur avisé.

Notre étude se propose de développer une méthodologie visant à intégrer les aspects liés à la mémorisation dans le flot de synthèse de l'outil Gaut Low Power. L'objectif est de trouver une solution architecturale répondant aux contraintes de l'application et à un compromis entre le temps de conception et la qualité de la solution.

## 3. Stratégie de conception

Nous nous intéressons aux applications de traitement du signal et de l'image ayant un comportement déterministe ; ces applications peuvent être modélisées par des graphes flots de données (DFG) et les accès mémoires sont connus à l'avance. La synthèse d'architecture de telles applications, caractérisées par des besoins de mémorisation importants, permet d'explorer de nombreuses alternatives d'implantation de hiérarchie mémoire. Cependant, parmi toutes ces alternatives, seules les architectures conduisant aux meilleures performances et à de faibles consommations doivent être retenues. Aussi, il est nécessaire de contraindre le processus de synthèse et de mettre en œuvre des méthodes afin de restreindre l'espace de conception et d'aboutir rapidement aux meilleures solutions.

La conception d'unité de mémorisation lors de la synthèse d'architecture est un problème complexe. Notre stratégie de synthèse mémoire, présentée sur la figure 1, peut se décomposer en trois étapes : en amont, au cœur, et en aval de la synthèse d'architecture.

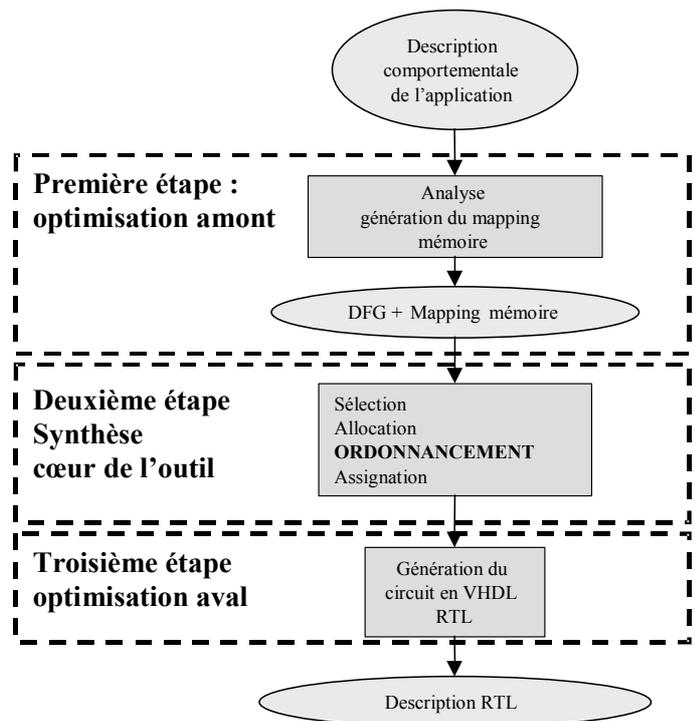

Figure 1 : stratégie de conception

Le point d'entrée est une description comportementale de l'application.

- La première étape se situe en amont de la synthèse d'architecture. Cette étape consiste en une analyse du code afin de modéliser l'application sous forme de DFG, permettant d'exprimer le parallélisme potentiel de l'application ainsi que les dépendances de données. L'analyse du code permet également d'extraire un mapping mémoire, à savoir définir une hiérarchie mémoire et de réaliser la distribution et le placement des structures (tableaux mono ou multidimensionnels) et des variables dans cette hiérarchie mémoire. L'objectif de l'étape amont est de réduire la consommation de la partie mémoire en minimisant le nombre de transferts et le nombre de transitions sur les bus d'adresses.

- L'étape de synthèse reprend les phases classiques de l'outil de synthèse d'architecture GAUT Low Power en y intégrant les contraintes liées au mapping mémoire déterminé dans la partie amont.

- Dans l'étape en aval de la synthèse, il nous faut définir une architecture pour chaque niveau de hiérarchie mémoire et les générateurs d'adresses.

Nous allons, dans une première approche, nous limiter au problème de l'intégration du mapping mémoire au cœur de la synthèse d'architecture. Nous supposons que la hiérarchie et le mapping mémoire sont réalisés ; nous développerons, dans la section suivante, les différentes étapes de synthèse de l'outil GAUT Low Power et plus précisément l'ordonnancement et ses modifications inhérentes à l'introduction d'un mapping mémoire.

## 4. La synthèse d'architecture

### 4.1 Particularités de GAUT Low Power

L'outil GAUT Low Power développé au LESTER permet, à partir de la description comportementale VHDL d'une application, d'obtenir une unité de traitement et une unité de contrôle au niveau RTL. La description RTL synthétisable est destinée aux outils de synthèse logique. Le but de cet outil est la synthèse d'architecture pour des applications en TDSI, orientées flot de données. Il intègre des critères de surface et consommation sous contrainte temps réel. Du point de vue de la consommation, GAUT Low Power prend en compte l'optimisation de la consommation tout au long des étapes du flot de synthèse (sélection, allocation, ordonnancement, assignation), notamment par la modélisation des signaux d'entrée des opérateurs (figure2).

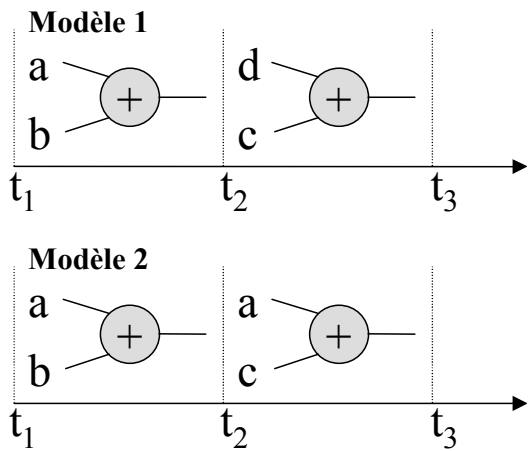

Figure 2 : modèles des signaux d'entrée des opérateurs

− Le modèle 1 : toutes les entrées de l'opérateur changent entre deux opérations consécutives de façon aléatoire. Cela représente le cas où la consommation liée à un opérateur est maximum.

− Le modèle 2 : une ou plusieurs entrées d'un opérateur sont identiques entre deux opérations successives, les autres variant aléatoirement. La consommation d'un tel opérateur est alors réduite de 25 à 50% par rapport au modèle 1 [2].

### 4.2 Ordonnancement dans GAUT Low Power

L'ordonnancement repose sur une technique classique par liste de priorités des opérations ; les opérations exécutables (Opr_exe) sont triées suivant leur mobilité. Le second critère permettant de choisir l'ordonnancement d'une opération repose sur le modèle des signaux d'entrée des opérateurs, présenté en figure 2. En effet, pour des opérations ayant la même priorité, les opérations satisfaisant le modèle 2 seront privilégiées. Le pseudo code de la figure 3 présente les différentes phases de l'ordonnancement. L'algorithme est réalisé, pour chaque date d'ordonnancement, tant qu'il y a des opérateurs libres (Opr_libre). Le choix de cet ordonnancement permet de réduire le taux d'activité global de la partie unité de traitement et la réduction de laconsommation peut atteindre 50% [2].

```
t = 0
Tant que (t < contrainte_temporelle)
    Pour Chaque tranche de pipeline
        Tant qu il existe Opr_libre
            Chercher Ops_exe
            Pour chaque Opr _libre
                Trier les Ops_exe (critère de mobilité)
                Choisir une Ops_exe (priorité au modèle 2)
                Ordonnancer cet Opr_libre
                Mettre à jour les listes des Opr et des Ops
            FinPour chaque Opr_libre
        FinTant que Opr_libre
    FinPour Chaque tranche de pipeline
Incrémenter t d un pas step
FinTant que (t < contrainte_temporelle)
```

Figure 3 : ordonnancement dans GAUT Low Power

L'absence de supposition sur la distribution et le placement des données lors de l'ordonnancement peut s'avérer néfaste pour l'optimisation de l'unité de mémorisation. Si toutes les opérations ordonnancées à un instant t ont besoin d'accéder à des variables placées en mémoire, alors le nombre important de bus, de bancs mémoire va entraîner une augmentation de la consommation de l'unité de mémorisation. Nous pouvons cependant tenter de réduire cette consommation en aval de la synthèse, mais les possibilités d'optimisation et les gains restent limités. Il faut donc traiter le problème de gestion du transfert et du placement des données à travers l'unité de mémorisation en amont de la synthèse et contraindre la synthèse de manière à limiter la consommation de l'unité de mémorisation.

### 4.3 Nouvel ordonnancement

Nous avons donc modifié cet ordonnancement. Les données en entrée des opérations peuvent provenir de mémoires ayant des temps d'accès différents et plusieurs opérations peuvent vouloir accéder à la même mémoire à un instant donné.

La figure 4 présente le nid de boucle modifié du nouvel ordonnancement dans lequel un nouveau critère est ajouté. La liste des opérations exécutables (Ops_exe) est triée suivant la mobilité des opérations. Les opérations exécutables ayant une même mobilité sont ensuite triées suivant le nombre d'entrées communes avec les opérations précédentes. Le choix de l'opération à ordonnancer s'effectue maintenant sur la possibilité ou non d'accéder aux données pour l'opération. Il faut, pour cela, ajouter un critère de conflit d'accès aux données placées dans l'unité de mémorisation. Pour effectuer ce choix, nous allons définir des opérateurs fictifs d'accès à chaque mémoire (Opr_acc_mem) dépendants du mapping mémoire. Un opérateur fictif d'accès, pour une mémoire donnée, sera associé à toutes les opérations sollicitant des données placées dans cette mémoire. L'opérateur fictif contient un champ indiquant s'il est libre ou occupé. Par exemple, si l'opérateur d'accès d'une mémoire est occupé, alors une opération exécutable nécessitant un accès à cette mémoire, bien que prioritaire dans la liste, ne pourra être ordonnancée. En résumé, chaque opération est associée à un ou plusieurs opérateurs fictifs d'accès mémoire. Si tous les opérateurs fictifs ne sont pas libres, alors l'opération ne peut être ordonnancée à l'instant t. Dans la liste triée des opérations exécutables, il faut sélectionner la première opération dont tous les opérateurs fictifs d'accès mémoire sont libres. L'opérateur libre est alors ordonnancé et assigné à l'opération sélectionnée. L'opérateur libre devient occupé (mise à jour des listes opérateurs_libres et opérateurs_occupés), l'opération commence à s'exécuter (mise à jour des listes opérations_exécutables et en_cours), les ports des différentes mémoires accédées sont occupés (mise à jours des opérateurs fictifs d'accès mémoire).

---

Pour chaque *Opr_libre*

   Trier liste des *Ops_exe* (premier critère : mobilité)

   Trier liste des *Ops_exe* (second critère : modèle 2 de conso)

   Sélectionner la 1ére Ops_exe dont ts les Opr_acc_mem sont libres

   Ordonnancer cet *Opr_libre*

   Mettre à jour les champs des *Opr_acc_mem*

   Mettre à jour les listes des *Opr* et des *Ops*

FinPour chaque *Opr_libre*

---

Figure 4 : nouvel ordonnancement

L'introduction de la prise en considération du mapping mémoire aura une influence sur les autres étapes de synthèse (sélection, allocation, assignation). Toutes ces étapes devront être également modifiées pour que l'outil GAUT Low Power puisse générer une description fonctionnelle de l'application au niveau RTL.

## 5. Perspectives

L'ordonnancement actuel de l'outil GAUT Low Power s'inscrit dans une politique d'optimisation de la partie opérative des algorithmes. Les différentes unités fonctionnelles sont fortement contraintes par l'unité de traitement ; de ce fait elles sont difficilement optimisables. La définition d'un mapping mémoire en amont de l'outil et la prise en compte de ce mapping au cœur de la synthèse va permettre de réaliser un compromis entre optimisation de l'unité de traitement et optimisation de l'unité de mémorisation. Le mapping mémoire va être intégré à toutes les phases de la synthèse ce qui permettra de déterminer son influence sur la consommation globale et de valider une méthode de détermination du mapping mémoire d'application en amont de la synthèse.

## 6. Bibliographie